\documentclass[fleqn,usenatbib]{mnras}

\usepackage[T1]{fontenc}
\usepackage{ae,aecompl}

\usepackage{graphicx}	% Including figure files
\usepackage{amsmath}	% Advanced maths commands
\usepackage{amssymb}	% Extra maths symbols
\usepackage{mathtools}
\usepackage{subfig}
\usepackage{tipa}
\usepackage[export]{adjustbox}
\usepackage{newtxtext,newtxmath}

\usepackage{soul}

\title[Lifetimes and Rotation in the SMMF]{Lifetimes and Rotation within the Solar Mean Magnetic Field}

\author[E. Ross et al.]{
Eddie Ross,$^{1,2}$\thanks{E-mail: exr007@student.bham.ac.uk (ER)}
William J. Chaplin,$^{1,2}$
Steven J. Hale,$^{1,2}$
Rachel Howe,$^{1,2}$
\newauthor{Yvonne P. Elsworth,$^{1,2}$, Guy R. Davies$^{1,2}$ and Martin Bo Nielsen$^{1,2,3}$}
\\
$^{1}$School of Physics and Astronomy, University of Birmingham, Edgbaston, Birmingham B15 2TT, UK\\
$^{2}$Stellar Astrophysics Centre, Department of Physics and Astronomy, Aarhus University, Ny Munkegade 120, DK-8000 Aarhus C,\\ Denmark\\
$^{3}$Center for Space Science, NYUAD Institute, New York University Abu Dhabi, PO Box 129188, Abu Dhabi, United Arab Emirates\\
}

\date{Accepted XXX. Received YYY; in original form ZZZ}

\pubyear{2020}

\begin{document}
\label{firstpage}
\pagerange{\pageref{firstpage}--\pageref{lastpage}}
\maketitle

\begin{abstract}
We have used very high-cadence (sub-minute) observations of the solar mean magnetic field (SMMF) from the Birmingham Solar Oscillations Network (BiSON) to investigate the morphology of the SMMF. The observations span a period from 1992--2012, and the high-cadence observations allowed the exploration of the power spectrum up to frequencies in the mHz range. The power spectrum contains several broad peaks from a rotationally-modulated (RM) component, whose linewidths allowed us to measure, for the first time, the lifetime of the RM source. There is an additional broadband, background component in the power spectrum which we have shown is an artefact of power aliasing due to the low fill of the data. The sidereal rotation period of the RM component was measured as $25.23 \pm 0.11$~days and suggests that the signal is sensitive to a time-averaged latitude of $\sim 12^{\circ}$. We have also shown the RM lifetime to be $139.6 \pm 18.5$~days. This provides evidence to suggest the RM component of the SMMF is connected to magnetic flux concentrations (MFCs) and active regions (ARs) of magnetic flux, based both on its lifetime and location on the solar disc.

\end{abstract}

\begin{keywords}
Sun: magnetic fields -- Sun: rotation -- Sun: activity
\end{keywords}

%%%%%%%%%%%%%%%%%%%%%%%%%%%%%%%%%%%%%%%%%%%%%%%%%%

%%%%%%%%%%%%%%%%% BODY OF PAPER %%%%%%%%%%%%%%%%%%

\section{Introduction}
\label{sec:intro}

The Sun has a complicated magnetic field structure; many features of the Sun and proxies for the solar activity are related to the evolution of the Sun's magnetic field. 

The solar mean magnetic field (SMMF) is a surprising, non-zero measurement of the imbalance of opposite polarities of magnetic flux observed on the full visible disc of the Sun \citep{svalgaard_suns_1975}, and is defined as the mean line-of-sight (LOS) magnetic field when observing the Sun-as-a-star \citep{scherrer_mean_1977, scherrer_mean_1977-1, garcia_integrated_1999}. In the literature the SMMF is also sometimes referred to as the general magnetic field (GMF) \citep{severny_time_1971} or the mean magnetic field (MMF) \citep{kotov_mean_2008} of the Sun.

Observations of the SMMF have typically been made by measuring the Zeeman splitting of spectral lines using a ground-based Babcock-type magnetograph \citep{scherrer_mean_1977}, although more recently the SMMF has been calculated from full-disc LOS magnetograms taken from space-borne telescopes such as the Solar Dynamics Observatory Helioseismic and Magnetic Imager (SDO/HMI) \citep{kutsenko_contribution_2017, bose_variability_2018}. It is understood that the strength of the SMMF may vary depending on the spectral line used to measure it \citep{kotov_mean_2008, kotov_enigmas_2012}; however, the SMMF varies slowly with the solar activity cycle, with an amplitude on the order of a Gauss during solar maximum and a tenth of a Gauss during solar minimum \citep{plachinda_general_2011}. In addition, the SMMF displays a strong, quasi-coherent rotational signal which must arise from inhomogeneities on the solar disc with lifetimes of several rotations \citep{chaplin_studies_2003, xie_temporal_2017}.

Despite existing literature on SMMF observations spanning several decades, the SMMF origin remains an open debate in solar physics. The principal component of the SMMF is commonly assumed to be weak, large-scale magnetic flux, distributed over the entire solar disc, rather than from magnetic flux concentrations (MFCs), active regions (ARs), or sunspots \citep{severny_time_1971, scherrer_mean_1977, xiang_ensemble_2016}. However, conversely, \citet{scherrer_mean_1972} found that the SMMF was most highly correlated with only the inner-most one quarter, by area, of the solar disc, which is more sensitive to active latitudes.

In recent literature, \citet{bose_variability_2018} provided a novel approach to understanding the SMMF whereby they decomposed the SMMF through feature identification and pixel-by-pixel analysis of full-disc magnetograms. They concluded that: (i) the observed variability in the SMMF lies in the polarity imbalance of large-scale magnetic field structures on the visible surface of the Sun; (ii) the correlation between the flux from sunspots and the SMMF is statistically insignificant; and (iii) more critically that the background flux dominates the SMMF, accounting for around 89 per cent of the variation in the SMMF. However, there still remained a strong manifestation of the rotation signal in the background component presented by \citet{bose_variability_2018}. This signal is indicative of inhomogeneous magnetic features with lifetimes on the order of several solar rotations, rather than the short-lived, weaker fields usually associated with the large-scale background. It therefore raises the question of whether their technique assigned flux from MFCs or ARs to the background. It is possible that some of the strong flux may have been assigned to the background signal, which then contributed to this rotation signal. 

Despite these findings, it is known that the strength of the SMMF is weaker during solar minimum, when there are fewer ARs, and stronger during solar maximum, when there are more ARs \citep{plachinda_general_2011}. This is suggestive that the evolution of ARs has relevance for the evolution of the SMMF.

There is a contrasting view in the literature which claims AR flux dominates the SMMF. \citet{kutsenko_contribution_2017} state that a large component of the SMMF may be explained by strong and intermediate flux regions. These regions are associated with ARs, suggesting between 65 to 95 per cent of the SMMF could be attributed to strong and intermediate flux from ARs, and the fraction of the occupied area varied between 2 to 6 per cent of the solar disc, depending on the chosen threshold for separating weak and strong flux. This finding suggests that strong, long-lived, inhomogeneous MFCs produce the strong rotation signal in the SMMF; however, \citet{kutsenko_contribution_2017} also discuss that there is an entanglement of strong flux (typically associated with ARs) and intermediate flux (typically associated with network fields and remains of decayed ARs). This means it is difficult to determine whether strong ARs or their remnants contribute to the SMMF. 

The Sun's dynamo and hence magnetic field is directly coupled to the solar rotation. The Sun exhibits latitude-dependent and depth-dependent differential rotation with a sidereal, equatorial period of around 25~days \citep{howe_solar_2009}. To Earth-based observers, the synodic rotation of the Sun is observed at around 27~days, and the SMMF displays a dominant signal at this period, and its harmonics \citep{chaplin_studies_2003, xie_temporal_2017, bose_variability_2018}. It was also reported by \citet{xie_temporal_2017} that the differential solar rotation was observed in the SMMF with measured synodic rotational periods of $28.28 \, \pm \, 0.67$~days and $27.32 \, \pm \, 0.64$~days for the rising and declining phases, respectively, of all of the solar cycles in their considered time-frame.

On the other hand, \citet{xiang_ensemble_2016} utilised ensemble empirical mode decomposition (EEMD) analysis to extract modes of the SMMF and found two rotation periods which are derived from different strengths of magnetic flux elements. They found that a rotation period of 26.6~days was related to weaker magnetic flux elements within the SMMF, while the measured period was slightly longer, at 28.5~days, for stronger magnetic flux elements.

In this work, we use high-cadence (sub-minute) observations of the SMMF, made by the Birmingham Solar Oscillations Network (BiSON) \citep{chaplin_bison_1996, chaplin_noise_2005, hale_performance_2016}, to investigate its morphology. This work provides a frequency domain analysis of the SMMF, and a rotationally-modulated (RM) component with a period of around 27~days is clearly observed as several peaks in the power spectrum. The breakdown of the paper is as follows. 

In Section~\ref{sec:data}, we provide an overview of the BiSON data used in this work; how the observations are made and the SMMF data are acquired. As this work provides an investigation of the SMMF in the frequency domain, in Section~\ref{sec:method} we discuss in detail how the power spectrum was modelled. 

In Section~\ref{sec:results} the results from modelling the power spectrum are presented. We outline the key findings and draw similarities between the properties of the RM component and ARs, suggesting that ARs may provide a strong contribution to the SMMF. Conclusions and discussions are presented in Section~\ref{sec:conc}. 
\section{Data}
\label{sec:data}

\subsection{Summary of the Data Set}

\citet{chaplin_studies_2003} provided the first examination of the SMMF using data from the Birmingham Solar Oscillations Network (BiSON), and the work presented in this paper is a continuation of that study. 

BiSON is a six-station, ground-based, global network of telescopes continuously monitoring the Sun, which principally makes precise measurements of the line-of-sight (LOS) velocity of the photosphere due to solar $p$ mode oscillations \citep{hale_performance_2016}. Through the use of polarizing optics and additional electronics, the BiSON spectrometers can measure both the disc-averaged LOS velocity and magnetic field in the photosphere \citep{chaplin_studies_2003}; however, not all BiSON sites measure the SMMF. 

In this study we focus on the data collected by the Sutherland node in South Africa, which was also used by \cite{chaplin_studies_2003}. This is the only station that has had the capability to measure and collect data on the SMMF long-term. Data are sampled on a 40-second cadence, and the SMMF data collected by the Sutherland station span the epoch from 01/1992 -- 12/2012 (i.e. covering 7643~days). Over this period, the average duty cycle of the 40-second data is $\sim 15.6$ per cent. If instead we take a daily average of the BiSON SMMF, the average duty cycle is $\sim 55.2$ per cent. This gives a higher duty cycle but a lower Nyquist frequency. Because of the much lower Nyquist frequency, modelling the background power spectral density is more challenging; therefore we use the 40-second cadence data in this work. However, both data sets return similar results; we discuss later in Section~\ref{sec:method_modelling} how we handled the impact of the low duty cycle of the 40-second data. A comparison of the daily-averaged SMMF observations made by BiSON to those made by the Wilcox Solar Observatory (WSO) is given in \citet{chaplin_studies_2003}.

\subsection{Obtaining the SMMF from BiSON}

To acquire the SMMF from BiSON data, the method described by \citet{chaplin_studies_2003} was adopted; here we discuss the key aspects.

Each BiSON site employs a resonant scattering spectrometer (RSS) to measure the Doppler shift of the Zeeman $^{2}\mathrm{S}_{1/2} \, \rightarrow \, ^{2}\mathrm{P}_{1/2}$ line of potassium, at 769.9~nm \citep{brookes_resonant-scattering_1978}. A potassium vapour cell placed within a longitudinal magnetic field Zeeman splits the laboratory line into the two allowed D1 transitions \citep{lund_spatial_2017}. The intensity of the longer wavelength (red; $I_R$) and shorter wavelength (blue; $I_B$) components of the line may be measured by the RSS almost simultaneously, by using polarizing optics to switch between the red and blue wings of the line, to form the ratio given by equation~(\ref{eq:ratio}) which is used as a proxy for the Doppler shift from the LOS velocity of the photosphere (see \citet{brookes_observation_1976, brookes_resonant-scattering_1978, elsworth_performance_1995, chaplin_studies_2003, broomhall_new_2009, davies_bison_2014, lund_spatial_2017}):

\begin{equation}
    \mathcal{R} = \frac{I_B - I_R}{I_B + I_R} \, .
	\label{eq:ratio}
\end{equation}

Photospheric magnetic fields Zeeman split the Fraunhofer line and the Zeeman-split components have opposite senses of circular polarization \citep{chaplin_studies_2003}. Additional polarizing optics are used in the RSS to manipulate the sense of circular polarization (either + or -) that is passed through the instrument. The ratio $\mathcal{R}_{+}$ or $\mathcal{R}_{-}$ is formed, and the ratios $\mathcal{R}_{\pm}$ would be equal if there was no magnetic field present.

\begin{figure}
	\includegraphics[width=\columnwidth]{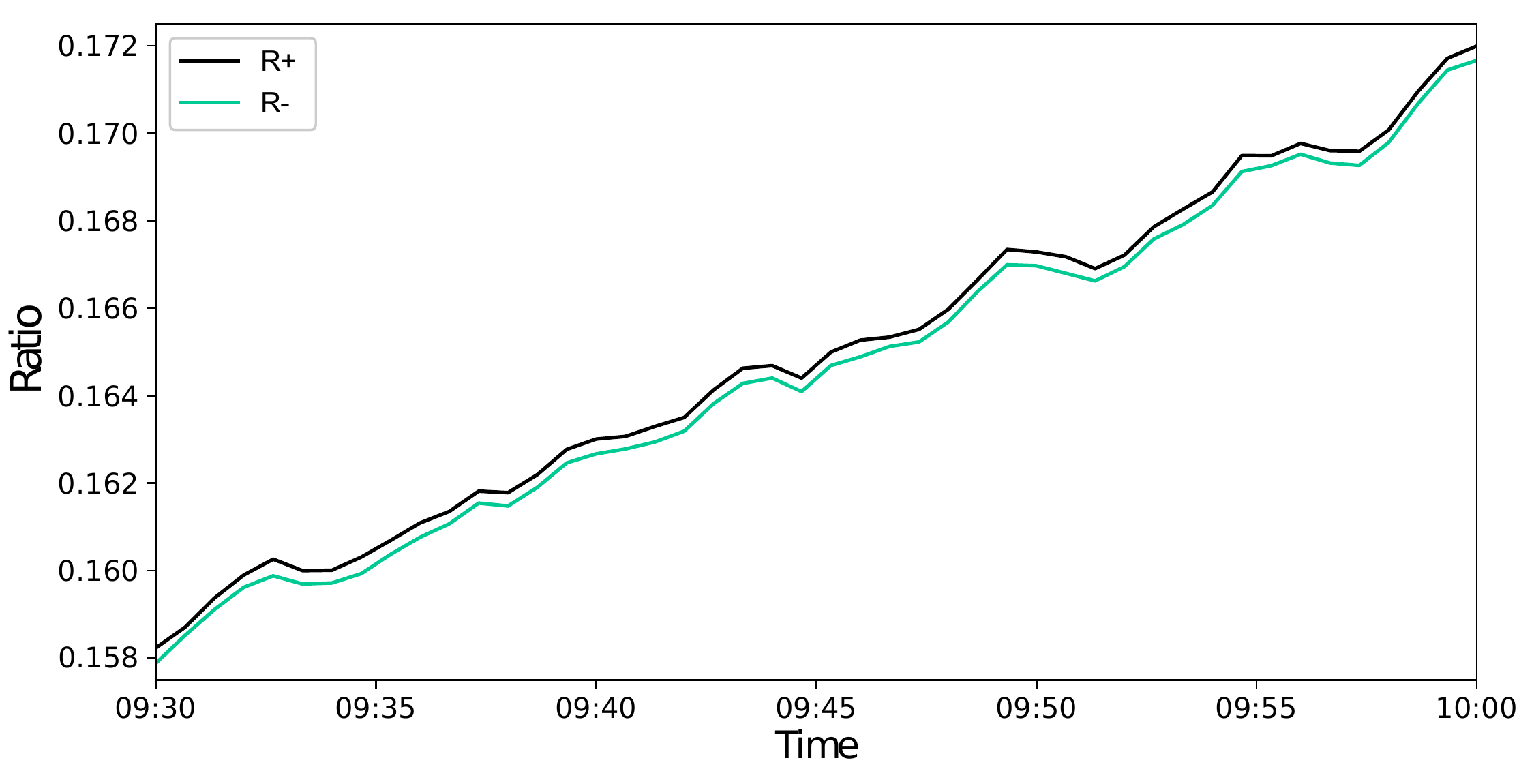}
    \caption{An example of the BiSON ratio data over a 30-minute period. The separation between the 2 ratios is due to the solar mean magnetic field and oscillations are due to the 5-minute $p$ mode signal.}
    \label{fig:ratio_split}
\end{figure}

The observed ratio ($\mathcal{R}_{\pm}$) may be decomposed as:

\begin{equation}
    \mathcal{R}_{\pm} = \mathcal{R}_{\mathrm{orb}} + \mathcal{R}_{\mathrm{spin}} + \mathcal{R}_{\mathrm{grs}} + \delta {r}_{\mathrm{osc}}(t) \pm \delta {r}_{\mathrm{B}}(t) \, ,
	\label{eq:vel_comp}
\end{equation}
where $\mathcal{R}_{\mathrm{orb}}$ is due to the radial component of the Earth's orbital velocity around the Sun, $\mathcal{R}_{\mathrm{spin}}$ is due to the component towards the Sun of the Earth's diurnal rotation about its spin axis as a function of latitude and time, $\mathcal{R}_{\mathrm{grs}}$ is from the gravitational red-shift of the solar line \citep{elsworth_techniques_1995, dumbill_observation_1999}, $\delta {r}_{\mathrm{osc}}(t)$ is due to the LOS velocity due to $p$ mode oscillations, and $\delta {r}_B(t)$ is due to the magnetic field ($\pm$ denotes the polarity of the Zeeman-split line that is being observed) \citep{dumbill_observation_1999}. The effect of the magnetic field on the ratio is shown in Fig.~\ref{fig:ratio_split}, and from equation~(\ref{eq:R_diff}), the difference between the opposite magnetic field ratios is twice the magnetic ratio residual, i.e.: 

\begin{equation}
    \mathcal{R}_{+} - \mathcal{R}_{-} = 2 \, \delta {r}_{\mathrm{B}}(t) \, .
	\label{eq:R_diff}
\end{equation}

The BiSON RSS is measuring the velocity variation on the solar disc, and therefore a calibration from the ratio to a velocity is necessary. One method of calibration is achieved by first fitting a 2nd- or 3rd-order polynomial as a function of velocity to the observed ratio averaged over both magnetic polarities, as discussed by \citet{elsworth_techniques_1995}. Here we chose to fit the ratio in terms of velocity, $\mathcal{R}_{\mathrm{calc}}(u)$, i.e.,

\begin{equation}
    \mathcal{R}_{\mathrm{calc}}(u) = \sum_{n} \mathcal{R}_{n} u^n \, ,
	\label{eq:calc_ratio}
\end{equation}
where:
\begin{equation}
    u = v_{\mathrm{orb}} + v_{\mathrm{spin}} \, ,
	\label{eq:stn_vel}
\end{equation}
and $v_{\mathrm{orb}}$ is the velocity component related to the ratio,  $\mathcal{R}_{\mathrm{orb}}$; $v_{\mathrm{spin}}$ is related to the ratio, $\mathcal{R}_{\mathrm{spin}}$; $n$ is the polynomial order.

It is possible to see that through the removal of $\mathcal{R}_{\mathrm{calc}}(u)$ (which we set up to also account for $\mathcal{R}_{\mathrm{grs}}$) from the observed ratios, one is left with the ratio residuals of the $p$ mode oscillations and the magnetic field, i.e.,

\begin{equation}
    \mathcal{R}_{\pm} - \mathcal{R}_{\mathrm{calc}}(u) = \delta {r}_{\mathrm{osc}}(t) \pm \delta {r}_{\mathrm{B}}(t) \, .
	\label{eq:ratio_resid}
\end{equation}

Furthermore, conversion from ratio residuals into velocity residuals uses the calibration given by equation~(\ref{eq:vel_resid}):

\begin{equation}
    \delta v(t) = \left( \frac{d\mathcal{R}_{calc}}{dV} \right)^{-1} \, \delta {r}(t)
	\label{eq:vel_resid} \, .
\end{equation}

\begin{figure}
\subfloat[Time-series of BiSON 40-s cadence SMMF  \label{fig:SMMF_TS}]{\includegraphics[width=0.98\columnwidth, right]{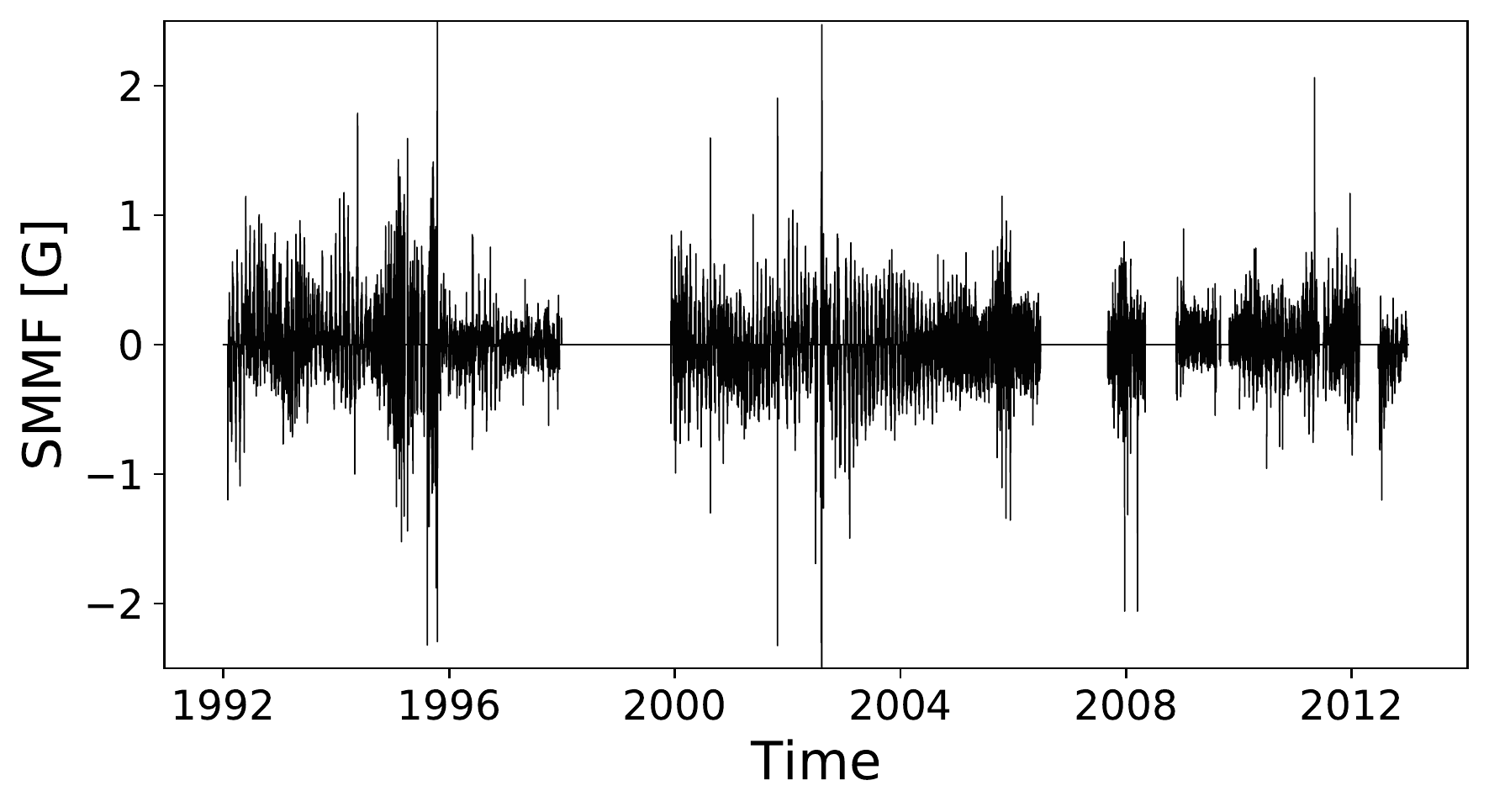}} \\ 
\subfloat[Power spectrum of BiSON 40-s cadence SMMF \label{fig:SMMF_FT}]{\includegraphics[width=\columnwidth]{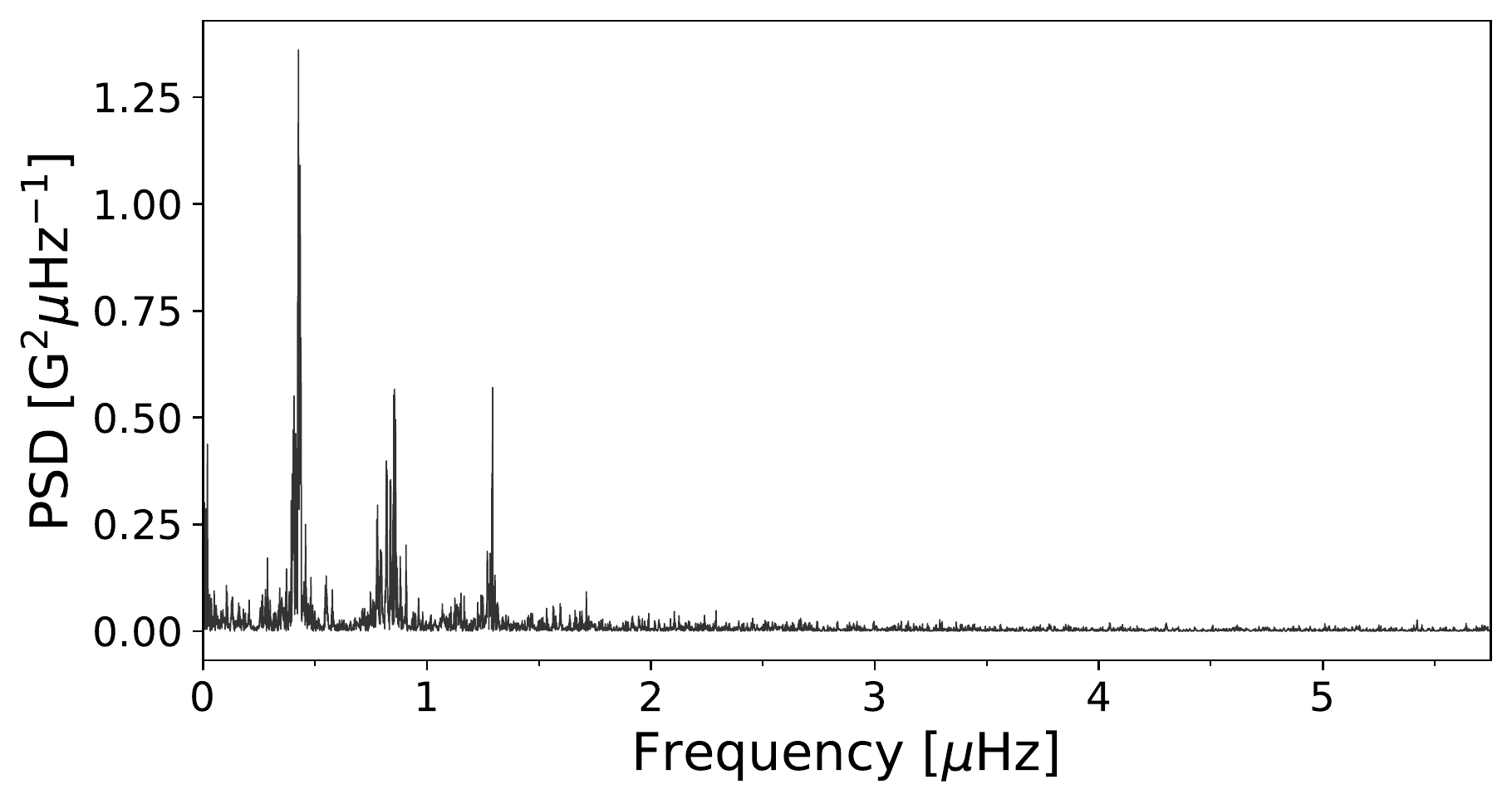}}
\caption{(a) 40-second cadence observations of the SMMF from the Sutherland BiSON station between 1992 and 2012. The sense of the field was chosen to match both \citet{chaplin_studies_2003} and the WSO observations, where positive is for a field pointing outwards from the Sun. (b) Power spectrum of the SMMF on a 40-second cadence truncated to $10 \, \mu\mathrm{Hz}$, however the Nyquist frequency is 12500~$\mu$Hz.}  \label{fig:BiSON_SMMF}
\end{figure}

In order to finally obtain the SMMF in units of magnetic field, one must combine equation~(\ref{eq:R_diff}) and  equation~(\ref{eq:vel_resid}) with the conversion factor in equation~(\ref{eq:K_B}) \citep{dumbill_observation_1999}, and the entire procedure can be simplified into:

\begin{equation}
    B(t) = \frac{1}{2} \left( \frac{d\mathcal{R}_{calc}}{dV} \right)^{-1} \frac{(\mathcal{R}_{+} - \mathcal{R}_{-})}{K_B} \, ,
	\label{eq:simplified_SMMF_cal}
\end{equation}
where,
\begin{equation}
    K_B = \frac{8}{3} \, \frac{\mu_B}{h} \, \frac{c}{\nu} \approx 2.89 \, \mathrm{ms}^{-1} \, \mathrm{G}^{-1} \, ,
	\label{eq:K_B}
\end{equation}
and $\mu_B$ is the Bohr magneton, $h$ is Planck's constant, $c$ is the speed of light, and $\nu$ is the frequency of the photons, 

Through the application of this methodology, one acquires the SMMF as shown in Fig.~(\ref{fig:SMMF_TS}). The power spectrum of the full, 7643-day Sutherland data set is shown in Fig.~(\ref{fig:SMMF_FT}), and it shows a strong rotational signal at a period of $\sim27$~days. The power spectrum of the SMMF is shown again in Fig.~\ref{fig:SMMF_40s_PSD} on a logarithmic scale covering the entire frequency range, which highlights the broadband background component of the power spectrum. 

\section{Methodology}
\label{sec:method}

\subsection{Parametrization of the SMMF Power Spectrum}
\label{sec:method_model_lifetimes}

\begin{figure}
	\includegraphics[width=\columnwidth]{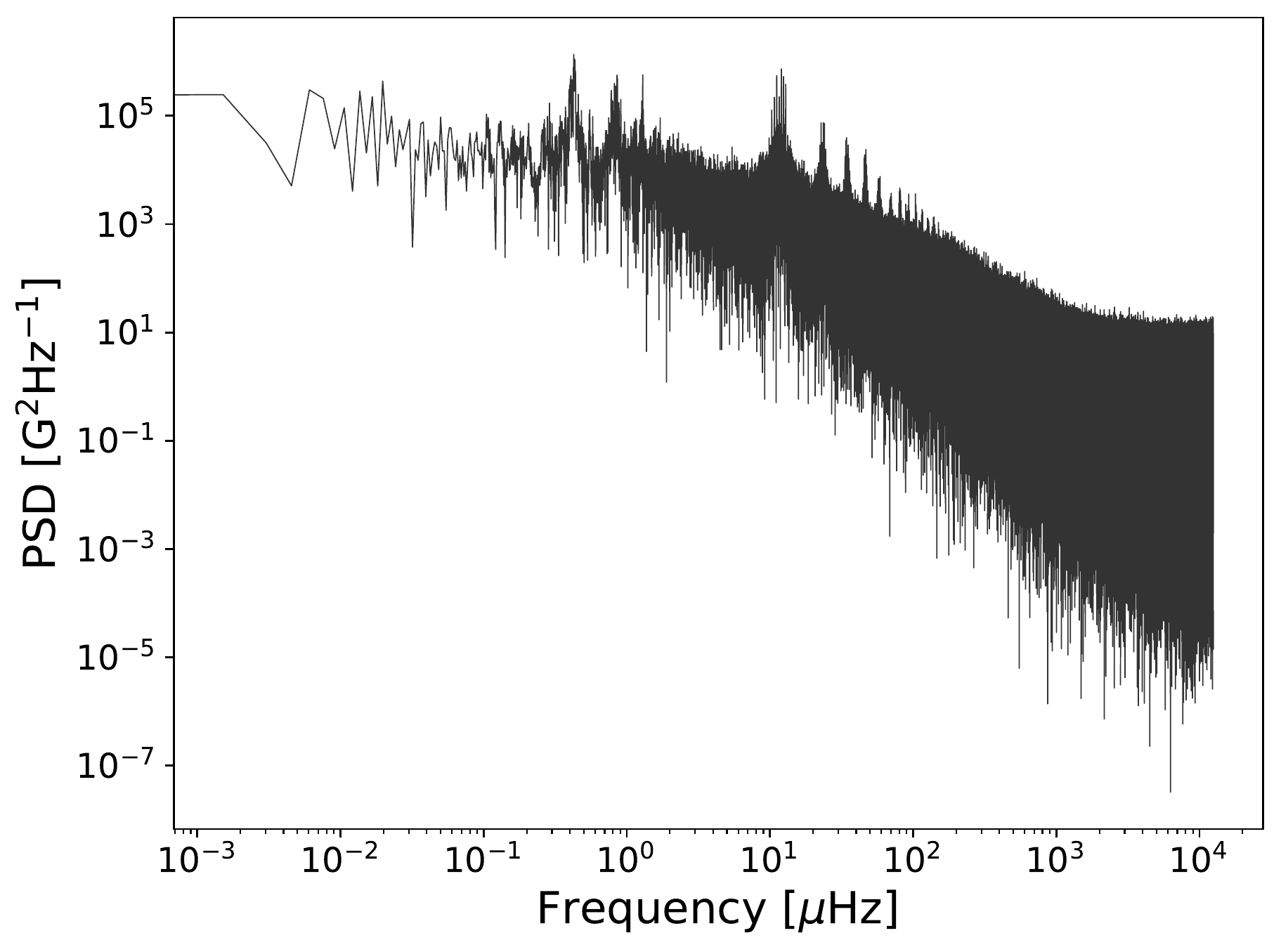}
    \caption{Power spectrum of 40-second cadence SMMF from the Sutherland BiSON station observed between 1992 -- 2012 on a logarithmic scale up to the full Nyquist frequency.}
    \label{fig:SMMF_40s_PSD}
\end{figure}

As we have 40-second cadence observations of the SMMF, we were able to investigate the power spectrum up to a Nyquist frequency of 12500~$\mu$Hz. There are a number of features within the full SMMF power spectrum, shown in Fig.~\ref{fig:SMMF_40s_PSD}. 

The peaks between 0.2 -- 2.0 $\mu\mathrm{Hz}$ in Fig.~\ref{fig:SMMF_FT} are a manifestation of rotation in the SMMF. The distinct set of peaks indicates the existence of a long-lived, inhomogeneous, rotationally-modulated (RM) source. Due to the quasi-coherent nature of the SMMF, and based on the comparatively short timescales for the emergence of magnetic features compared to their slow decay \citep{zwaan_solar_1981, harvey_properties_1993, hathaway_sunspot_2008}, we assume the evolution of individual features that contribute to the RM component may be modelled by a sudden appearance and a long, exponential decay. In the frequency-domain, each of the RM peaks may therefore be described by a Lorentzian distribution:

\begin{equation}
L_n(\nu; \Gamma, A_n, \nu_n) = \frac{2{A_n}^2}{\pi \Gamma} \left(1 + \left(\frac{(\nu - \nu_n)}{\Gamma/2} \right)^2\right)^{-1} \, ,
\label{eq:symm_lorentzian}
\end{equation}
where $\nu$ is frequency, $A_n$ is the root-mean-square amplitude of the peak, $\Gamma$ is the linewidth of the peak, $\nu_n$ is the frequency of the peak, and $n$ simply flags each peak. The mean-squared power in the time domain from the RM component of the SMMF is given by the sum of the ${A_n}^2$ of the individual harmonics in the power spectrum. 

Through this formulation we can measure the $e$-folding time ($T_e$) of the amplitude of the RM component, as it is related to the linewidth of the peak by:
\begin{equation}
\Gamma  = (\pi \, T_e)^{-1} \, .
\label{eq:mode_lifetime}
\end{equation}

The low-frequency power due to instrumental drifts, solar activity, and the window function can be incorporated into the model via the inclusion of a zero-frequency centred Lorentzian \citep{basu_asteroseismic_2017}, given by: 
\begin{equation}
H(\nu; \sigma, \tau) = \frac{4{\sigma}^2\tau}{1 + (2\pi \nu\tau)^2} \, ,
\label{eq:harvey}
\end{equation}
where $\sigma$ is the characteristic amplitude of the low frequency signal, and $\tau$ describes the characteristic timescale of the excursions around zero in the time-domain.

Finally, the high frequency power is accounted for by the inclusion of a constant offset due to shot-noise, $c$ \citep{basu_asteroseismic_2017}.

In the absence of any gaps in the data, the model function used to describe the power spectrum is given by: 

\begin{equation}
    P(\nu, \,{\bf a}) = \sum_{n=1}^{N} L_n(\nu; \Gamma, A_n, \nu_n) \, + \, H(\nu; \sigma, \tau) \, + \, c \, ;
    \label{eq:PSD_fit}
\end{equation}
the subscript, $n$, describes a single peak in the power spectrum. In implementing the model we constrain the mode frequencies such that they must be integer values of $\nu_0$: $\nu_n \, = \, n \nu_0$. This means that we define a single rotation frequency only, $\nu_0$, and subsequent peaks are the harmonic frequencies. It is worth noting explicitly that this function assumes the line width of each Lorentzian peak is the same, only the amplitudes and central frequencies differ.

The duty cycle of the Sutherland SMMF observations is very low, $\sim 15$ per cent, therefore it was important to take into consideration the effect that gaps in the data have on the power spectrum. Gaps in the data cause an aliasing of power from the signal frequencies to other frequencies in the spectrum, and the nature of the aliasing depends on the properties of the window function of the observations.

\begin{figure}
    \centering
	\includegraphics[width=\columnwidth]{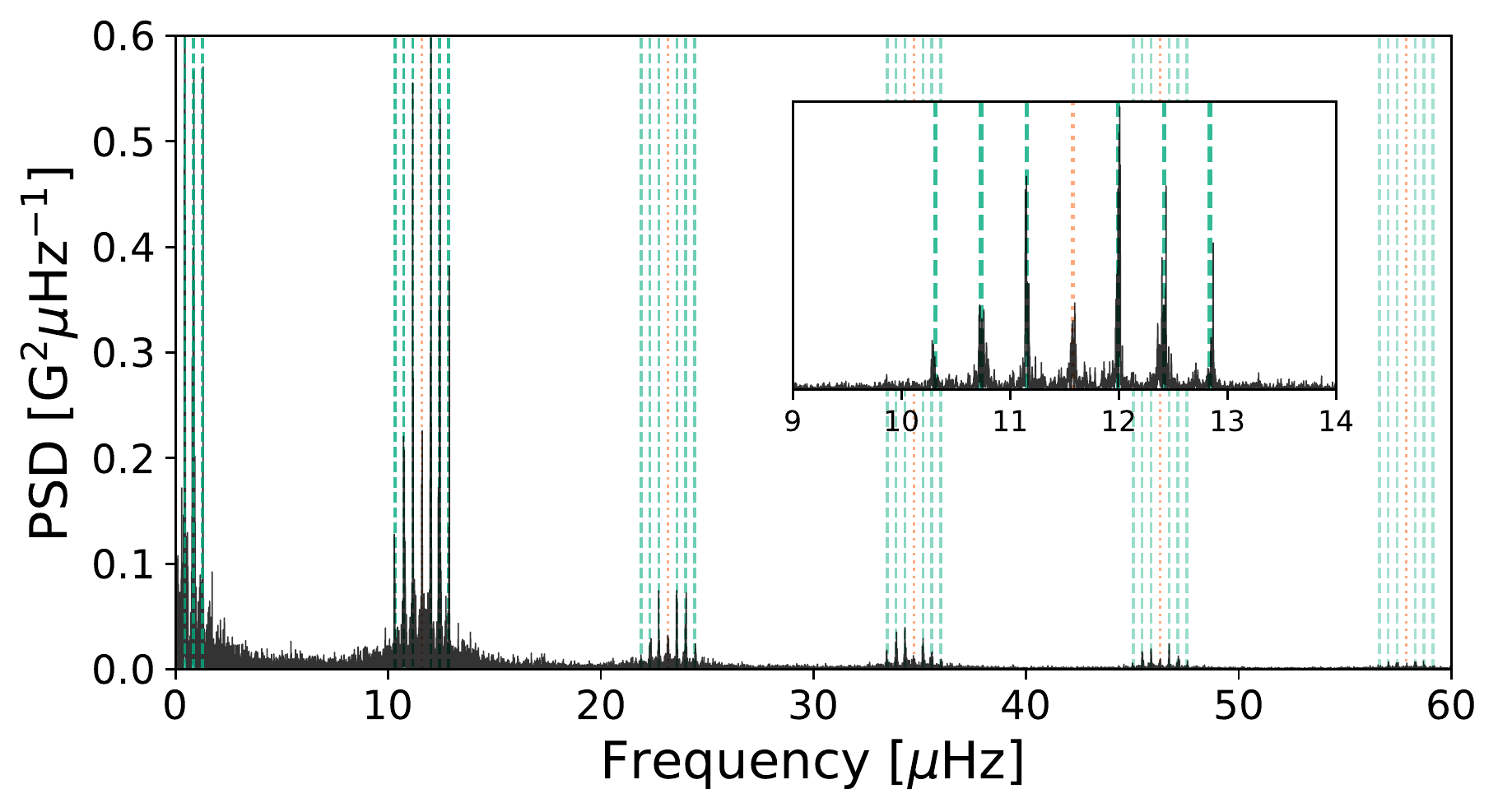}
    \caption{Locations of aliased power in sideband peaks. The orange dotted-lines show the locations of frequencies at multiples of 1/day. The green dashed-lines show the location of the sideband peaks -- harmonic frequencies reflected around multiples of 1/day.  The inset shows a zoom of one set of sideband peaks around 1/day.}
    \label{fig:sidebands}
\end{figure}

Periodic gaps in the data give rise to sidebands in the power spectrum and random gaps cause a more broadband shifting of power, meaning that some power from the low-frequency RM component is aliased to higher frequencies. The daily, periodic gaps in the BiSON data, due to single-site observations, produce sidebands around a frequency of 1/day, i.e. $\sim$~11.57 $\mu$Hz, and its harmonics. The aliased power is therefore located at frequencies:
\begin{equation}
    \nu_{n, i} = i \, (\frac{1}{\mathrm{day}} \pm \nu_{n}) \, , 
    \label{eq:sidebands}
\end{equation}
where $i$ denotes the sideband number and $n$ denotes the harmonic of the mode. The sideband structure implied by equation~(\ref{eq:sidebands}) is shown clearly in Fig.~\ref{fig:sidebands}.

The tails of the aliased peaks are long, therefore aliased power was re-distributed across the entire frequency range which produced a red-noise-like background component. To understand the broadband effects of the window function we generated an artificial time series from a single Lorentzian (representing the fundamental RM component). The artificial data were generated by calculating the inverse Fourier transform of the power spectrum which had the same Nyquist frequency and frequency resolution of the SMMF power spectrum. We then injected the gaps from the BiSON observations into this artificial time series, to ensure the window function was the same as the BiSON SMMF, and finally investigated the resultant power spectrum both without and without the window function.

\begin{figure}
	\includegraphics[width=\columnwidth]{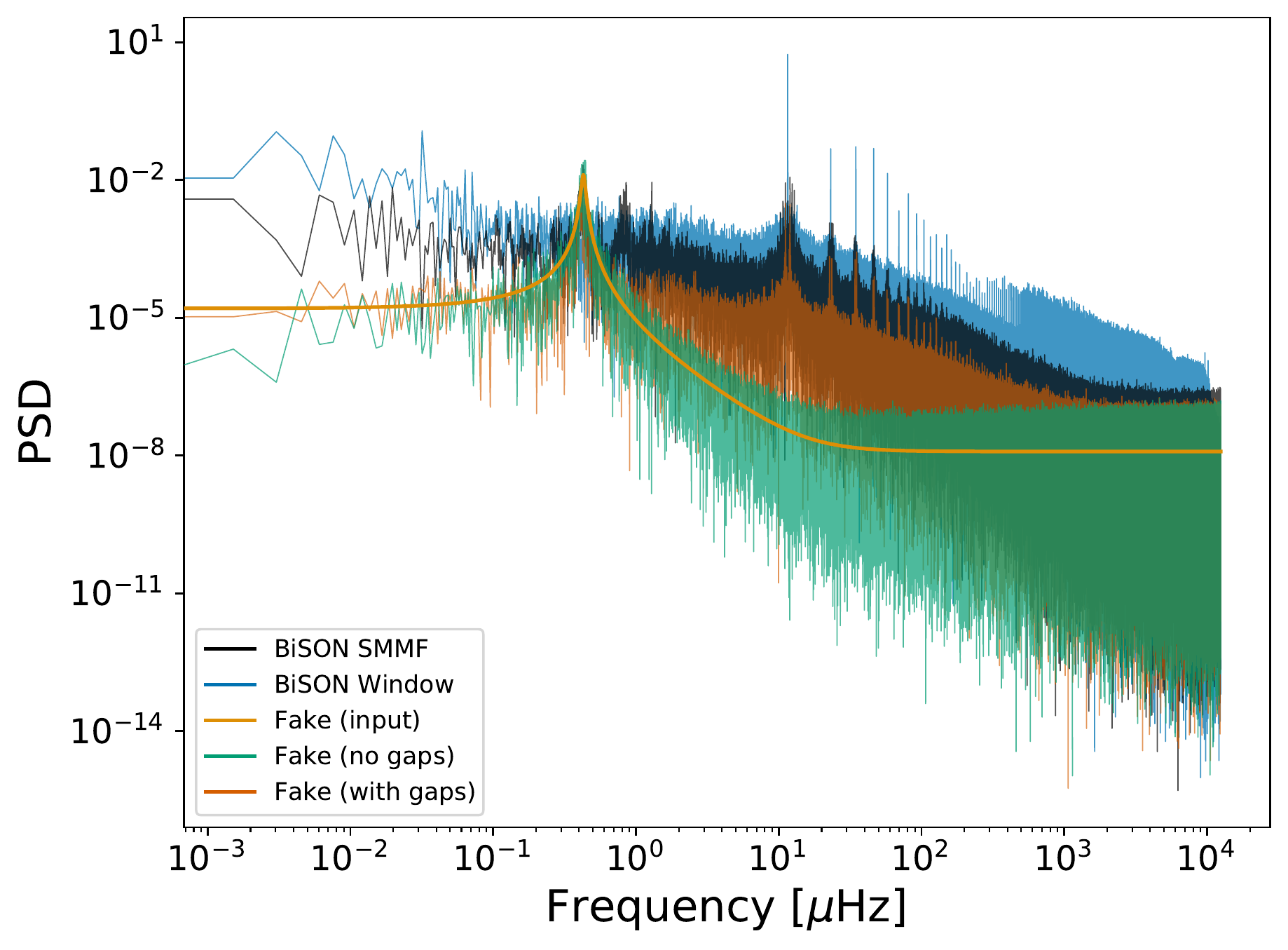}
    \caption{The effects of the window function on the power spectrum is shown by using a fake data set and this is compared to the BiSON power spectrum. Black line: BiSON SMMF power spectrum; blue line: power spectrum of the window function; green and dark-orange lines: the power spectrum of the artificial data without and with gaps, respectively; amber line: the input peak used to generate the artificial data over-plotted. The power spectra of the BiSON SMMF and the window function have been shifted upwards by a factor of 6 and 30, respectively, for clarity.}
    \label{fig:PSDs}
\end{figure}

Fig.~\ref{fig:PSDs} shows the effect of the window function on the resultant power spectrum. The power spectrum generated from the time series without gaps produces a single Lorentzian peak (amber and green lines). The injection of gaps into the time series (orange line) produces both the red-noise-like background component, as well as the sidebands, which bears a striking resemblance to the power spectrum of the BiSON SMMF observations (black line) and also the power spectrum of the window function (blue line).

This shows that the BiSON SMMF spectrum has a red-noise-like background component that is not due to any ephemeral signal, but due to the re-distribution of power by the window function of the BiSON observations.

In the time domain, the observed data, $y(t)$, includes the window function which, analytically, we can express as a multiplication of the uninterrupted, underlying signal, $f(t)$, with the window function, $g(t)$:
\begin{equation}
y(t)  = f(t) \, g(t)
\label{eq:timeseries} \, ,
\end{equation}
where: 
\begin{equation}
    g(t) = 
  \begin{cases} 
   1 & \text{for } |B(t)| > 0 \\
   0       & \text{for } |B(t)| = 0
  \end{cases} \, .
    \label{eq:window}
\end{equation}

Multiplication in the time domain becomes a convolution in the frequency domain. To model the observed power spectrum in a robust manner, taking into account the intricacies caused by gaps in the data, we used a model which was formed of a model power spectrum, $P(\nu; \,{\bf a})$ (equation~(\ref{eq:PSD_fit})), convolved with the Fourier transform of the window function of the observations ($\left|\mathcal{F}\left[g(t)\right]\right|^2$), i.e.,

\begin{equation}
    P'(\nu, \,{\bf a}) \, = \, P(\nu, \,{\bf a}) * \left|\mathcal{F}\left[g(t)\right]\right|^2 \, .
    \label{eq:PSD_fit_conv}
\end{equation}

Care was taken during this operation to ensure Parseval's theorem was obeyed, that no power was lost or gained from the convolution:

\begin{equation}
    \sum_{\nu} P'(\nu) \, = \,\sum_{\nu} P(\nu) \, = \, \frac{1}{N} \sum_{t}  B(t)^2 \, ,
    \label{eq:parseval}
\end{equation}
where $N$ here is the number of observed cadences.

\subsection{Modelling the SMMF Power Spectrum}
\label{sec:method_modelling}

Parameter estimation using the model defined in the previous section,  including all parameters, ${\bf a}$, was performed in a Bayesian manner using a Markov Chain Monte Carlo (MCMC) fitting routine.

Following from Bayes’ theorem we can state that the posterior probability distribution, $p({\bf a} | D, I)$, is proportional to the likelihood function, $L(D | {\bf a}, I)$, multiplied by a prior probability distribution, $p({\bf a} | I)$:

\begin{equation}
    p({\bf a} | D, I) \propto L(D | {\bf a}, I) \, p({\bf a} | I) \, ,
    \label{eq:bayes}
\end{equation}
where $D$ are the data, and $I$ is any prior information.

To perform the MCMC integration over the parameter space we must define a likelihood function; however, in practice, it is more convenient to work with logarithmic probabilities. The noise in the power spectrum is distributed as $\chi^2$ 2 degrees-of-freedom \citep{anderson_modeling_1990, handberg_bayesian_2011, davies_low-frequency_2014}, therefore the log likelihood function is:

\begin{equation}
    \ln{(L)} = - \sum\limits_{i} \left\{ \ln{(M_{i}({\bf a}))} + \frac{O_i}{M_{i}({\bf a})} \right\} \, ,
    \label{eq:likelihood_functino}
\end{equation}
for a model, $M_i$, with parameters, ${\bf a}$, and observed power, $O_i$, where $i$ describes the frequency bin. This likelihood function assumes that all the frequency bins are statistically independent but the effect of the window function means that they are not. We handled this issue in the manner described below, which used simulations based on the artificial data discussed in Section~\ref{sec:method_model_lifetimes}.

The prior information on each of the parameters used during the MCMC sampling were uniform distributions (denoted by $\mathcal{U}(l, u)$ with $l$ and $u$ representing the lower and upper limits of the distribution, respectively):

\begin{gather*}
\nu_0 \, \sim \,\mathcal{U}(0.38, 0.50) \> \mu\mathrm{Hz} \\
\Gamma \, \sim \,\mathcal{U}(0.00, 0.11)  \> \mu\mathrm{Hz} \\
A_1 \, \sim \,\mathcal{U}(100, 350)  \> \mathrm{mG} \\
A_2 \, \sim \,\mathcal{U}(50, 200)  \> \mathrm{mG} \\
A_3 \, \sim \,\mathcal{U}(20, 150)  \> \mathrm{mG} \\
A_4 \, \sim \,\mathcal{U}(10, 100)  \> \mathrm{mG} \\
\sigma \, \sim \,\mathcal{U}(0.01, 500)  \> \mathrm{mG} \\
\tau \, \sim \,\mathcal{U}(0.10, 200)  \> 10^6 \, \mathrm{s} \\
c \, \sim \,\mathcal{U}(10^{-3}, 10^{2})  \> \mathrm{G}^2 \, \mathrm{Hz}^{-1} \, . \\
\end{gather*}

The limits on the priors were set to cover a sensible range in parameter space, whilst limiting non-physical results or frequency aliasing.

The power spectrum of the 40-second cadence SMMF was modelled using equation~(\ref{eq:PSD_fit_conv}) (with $N = 4$ Lorentzian peaks in $P(\nu, \,{\bf a})$) using the affine-invariant MCMC sampler \textsc{emcee} \citep{foreman-mackey_emcee_2013} to explore the posterior parameter space.

The chains are not independent when using \textsc{emcee}, therefore convergence was interrogated using the integrated autocorrelation time. We computed the autocorrelation time using \textsc{emcee} and found $\tau \sim 120$~steps. \cite{foreman-mackey_emcee_2013} suggests that chains of length $\geq 50\tau$ are often sufficient. After a burn in of 6000 steps, we used 7000 iterations on 50 chains to explore the posterior parameter space, which was sufficient to ensure we had convergence on the posterior probability distribution.

As a result of the convolution in the model the widths of the posterior distributions for the model parameters were systematically underestimated. This effect arises because we do not account explicitly for the impact of the window function convolution on the covariance of the data; it is difficult to overcome computationally, especially with such a large data set ($\sim 10^7$ data points). To overcome this issue we performed the simulations using artificial data, described above, both with and without the effects of the window function and the use of the convolution in the model. This helped us to understand how the convolution affected our ability to measure the true posterior widths, which allowed us to account for the systematic underestimate of the credible regions of the posterior when modelling the power spectrum of the observed BiSON SMMF.

We also analysed the data as daily, one-day-cadence averages; this gave a higher fill ($\sim 55$ per cent) but a lower Nyquist frequency ($\sim 5.787$~mHz). Because of the much lower Nyquist, modelling the background power spectral density was more challenging but the duty cycle was approximately three times higher, resulting in a smaller effect from the window function. We note that we recovered results in our analysis of the daily averaged data that were consistent with those from the analysis of the data with a 40-second cadence.

\section{Results}
\label{sec:results}

\subsection{Rotation}
\label{sec:rot_results}

From the adjusted posterior distributions for each of the parameters, acquired through modelling the power spectrum, we were able to measure the fundamental rotational frequency and linewidth of the RM component. The latter was assumed to be the same for each peak.

\begin{figure}
\subfloat[Full power spectrum of the SMMF on logarithmic axes  \label{fig:full_PSD_fit}]{\includegraphics[width=0.98\columnwidth, right]{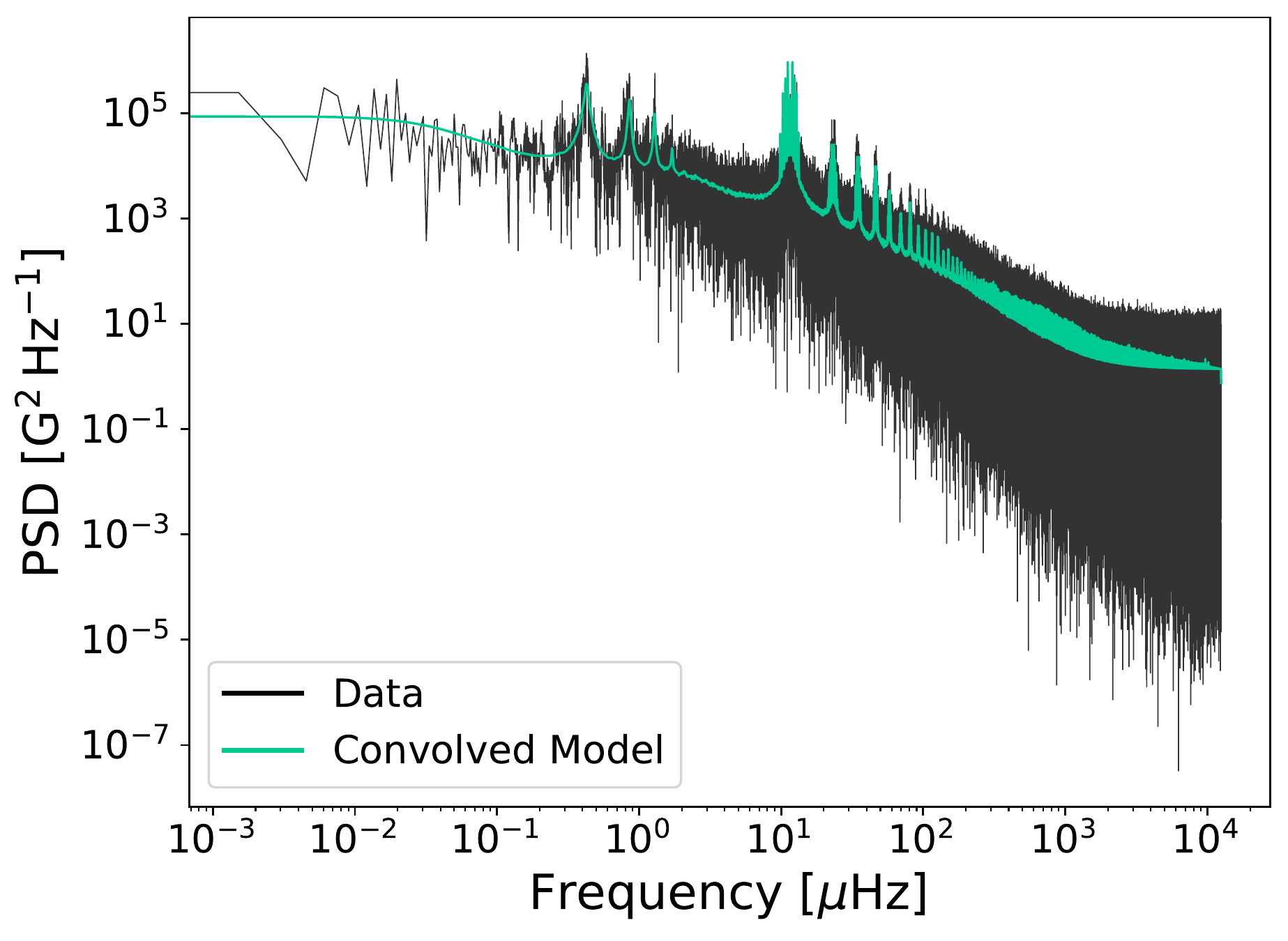}} \\ 
\subfloat[Power spectrum of the SMMF on linear axes, up to a frequency of $2.5 \, \mu\mathrm{Hz}$ \label{fig:full_PSD_fit_linear}]{\includegraphics[width=\columnwidth]{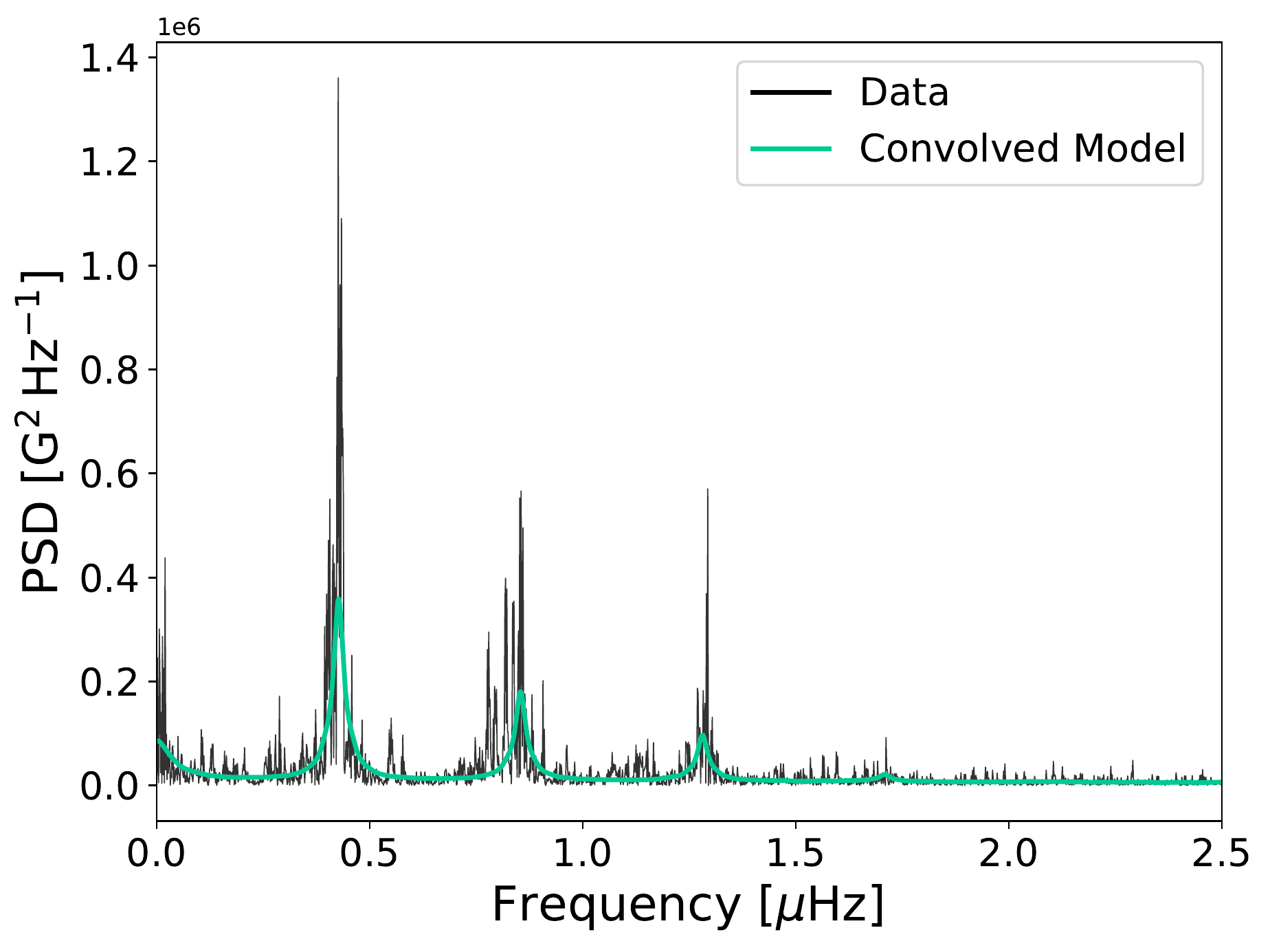}}
\caption{Power spectrum and the best-fitting model for: (a) the full power spectrum of the SMMF on logarithmic axes. (b) Power spectrum of the SMMF on linear axes, up to a frequency of $2.5 \, \mu\mathrm{Hz}$ in order to show the fundamental signal peaks due to rotation-modulated ARs. The data is displayed in black and the model is shown in green.}  \label{fig:PSD_fits}
\end{figure}

In Table~\ref{tab:full_fit_params} the median values of marginalised posterior distributions for each of the model parameters of equation~(\ref{eq:PSD_fit_conv}) are displayed. The resultant posterior distributions were approximately normally distributed and there was no significant covariance between parameters, therefore reported uncertainties on the parameters correspond to the $68$ per cent credible intervals either side of the median in the posterior distributions, adjusted for the systematic window function effects. In addition, we show the raw data with the model fit over-plotted in Fig.~\ref{fig:full_PSD_fit} and Fig.~\ref{fig:full_PSD_fit_linear}, on logarithmic and linear scales, respectively, to highlight the fit over the full frequency range, and the RM peaks, respectively.

\begin{table}
%\centering
\caption{Power spectrum model median results. Numbers in brackets
denote uncertainties on the last 2 digits, and all uncertainties correspond to the $68 \%$ credible intervals either side of the median for adjusted posterior widths.}
\label{tab:full_fit_params}
\begin{tabular}{l r r l r r }
    \hline
    {\bf $\theta$}  &  {Value}  & {Unit} &  {\bf $\theta$}  &  {Value}  & {Unit} \\
    \hline
	{$\nu_0$}  &  {0.4270$\left(_{-18}^{+18}\right)$} & {$\mu\mathrm{Hz} $}  &  {$A_4$}  &  {32.6$\pm2.1$} & {$\mathrm{mG}$}	\\
	
    {$\Gamma$}  &  {0.0264$\left(_{-35}^{+35}\right)$} & {$\mu\mathrm{Hz} $}	&  {$\tau$}	 &  {51.8$\pm6.8$} & {$\mathrm{days}$}	\\
    
    {$A_1$}  &  {166.0$\pm10.7 $} & {$\mathrm{mG}$}	&  {$\sigma$}  &  {83.4$\pm5.4$} & {$\mathrm{mG}$}	\\
    
    {$A_2$}  &  {115.9$\pm7.4$} & {$\mathrm{mG}$}	&  {$c$}  &  {0.2103$\left(_{-03}^{+03}\right)$} & {$\mathrm{G}^2\mathrm{Hz}^{-1}$}	\\ 
    
    {$A_3$}  &  {83.2$\pm5.3$} & {$\mathrm{mG}$} & {}  & {} & {} \\ \hline
    
\end{tabular}
\end{table}

%%%

%%%

The central frequency of the model, $\nu_0$, implies a fundamental synodic rotation period of $27.11 \pm 0.11$~days, and hence a sidereal rotation period of $25.23\pm0.11$~days. The rotation period measured here is in agreement with other literature for the rotation signal in the SMMF \citep{chaplin_studies_2003, xie_temporal_2017}. 

According to the model for differential rotation in equation~(\ref{eq:diff_rot_freq}), the measured rotation period suggests that the observed SMMF is sensitive to a time-averaged latitude of around $12^{\circ}$. This latitude is consistent with those spanned by sunspots and ARs over the solar activity cycle \citep{maunder_note_1904, mcintosh_deciphering_2014}, and particularly during the declining phase of the solar cycle \citep{thomas_asteroseismic_2019}. This suggests the origin of the RM component of the SMMF could be linked to active regions.

\subsection{Lifetimes}

From the measured linewidth of the Lorentzian peaks, we have calculated the lifetime of the RM component using equation~(\ref{eq:mode_lifetime}). The linewidth suggests a RM lifetime of $139.6 \pm 18.5$~days, which is in the region of $\sim 20 \pm 3$~weeks. The effects of differential rotation and AR  migration do not impact our ability to measure the linewidth, and thus lifetime, of the peaks (as explained in Appendix~\ref{sec:smearing}).

The typical lifetime of active magnetic regions and sunspots is on the order of weeks to months \citep{zwaan_solar_1981, howard_sunspot_2001, hathaway_sunspot_2008}, therefore the observations of the SMMF by BiSON measure a lifetime of the RM component which is consistent with the lifetime of ARs and sunspots. This again suggests that the RM signal is linked to active regions of magnetic field, suggesting them as a possible source of the signal.

When verifying these results by repeating the analysis with a daily averaged SMMF (see Section~\ref{sec:method_modelling}), the results for the linewidth were consistent.

\section{Discussions and Conclusions}
\label{sec:conc}

We have presented, for the first time, a frequency-domain analysis of $\sim$20 years of high-cadence (40-second) BiSON observations of the SMMF.

The investigation of very high-cadence observations of the SMMF allowed the exploration of the power spectrum up to $12.5$~mHz and the long duration of observations provided near-nHz resolution in the power spectrum which allowed us to measure the parameters associated with the rotationally modulated (RM) component of the SMMF.

We have measured the central frequency of the RM component, allowing us to infer the sidereal period of the RM to be $25.23~\pm~0.11$~days. This rotation period matches to an activity cycle average latitude of $\sim 12^{\circ}$, which is in the region of the typical latitudes for active magnetic regions averaged over the activity cycle \citep{maunder_note_1904, mcintosh_deciphering_2014, thomas_asteroseismic_2019}.

For the first time, using the linewidth of the peaks we have measured the lifetime of the RM component in the SMMF. The lifetime of the source of the RM component was inferred to be $139.6~\pm~18.5$ days. This lifetime is consistent with those of active magnetic regions and sunspots, in the region of weeks to months \citep{zwaan_solar_1981, hathaway_sunspot_2008}. 

There has been considerable debate in the literature concerning the origin of the SMMF. In this study, as the properties of the RM component are consistent with ARs, we have presented novel evidence suggesting them as the source of the SMMF.

\section*{Acknowledgements}

We would like to thank all those who are, or have been, associated with BiSON. The authors would like to acknowledge the support of the UK Science and Technology Facilities Council (STFC). Funding for the Stellar Astrophysics Centre (SAC) is provided by The Danish National Research Foundation (Grant DNRF106). This research also made use of the open-source Python packages: \textsc{Astropy},\footnote{http://www.astropy.org} a community-developed core Python package for Astronomy \citep{robitaille_astropy_2013, the_astropy_collaboration_astropy_2018}, \textsc{corner} \citep{foreman-mackey_corner.py_2016}, \textsc{emcee} \citep{foreman-mackey_emcee_2013}, \textsc{Matplotlib} \citep{hunter_matplotlib_2007}, \textsc{Numpy} \citep{harris_array_2020}, \textsc{Pandas} \citep{mckinney_data_2010}, and \textsc{SciPy} \citep{jones_scipy_2001}.

\section*{Data Availability}

This work uses data from the Birmingham Solar-Oscillations Network (BiSON), which may be accessed via the \href{http://bison.ph.bham.ac.uk/opendata}{BiSON Open Data Portal}.\footnote{http://bison.ph.bham.ac.uk/opendata}

%%%%%%%%%%%%%%%%%%%%%%%%%%%%%%%%%%%%%%%%%%%%%%%%%%



\bibliographystyle{mnras}
\bibliography{paper} % if your bibtex file is called example.bib

%%%%%%%%%%%%%%%%%%%%%%%%%%%%%%%%%%%%%%%%%%%%%%%%%%

%%%%%%%%%%%%%%%%% APPENDICES %%%%%%%%%%%%%%%%%%%%%

\appendix
\section{Testing the Effects of Differential Rotation and Migration}
\label{sec:smearing}

As a result of solar differential rotation and the migration of ARs towards the solar equator during the activity cycle, it is understood that the rotation period of ARs will vary throughout the solar cycle. 

As we have inferred that the RM component of the SMMF is likely linked to ARs, we may therefore assume that the RM component is also sensitive to latitudinal migration. Here we analysed the effect of this migration and differential rotation on our ability to make inferences on the lifetime of the RM component.

Several studies have modelled the the solar differential rotation, and its variation with latitude and radius of the Sun (see \citet{beck_comparison_2000} and \cite{howe_solar_2009} for in-depth reviews of the literature on solar differential rotation). Magnetic features have been shown to be sensitive to rotation deeper than the photosphere; therefore in general magnetic features can be seen to rotate with a shorter period than the surface plasma \citep{howe_solar_2009}.

\citet{chaplin_distortion_2008} analysed the effects of differential rotation on the shape of asteroseismic $p$ modes of oscillation with a low angular degree (i.e. $l \leq 3$), and showed that the consequence of differential rotation is to broaden the observed linewidth of a mode peak. The authors provide a model of the resultant profile of a $p$ mode whose frequency is shifted in time to be a time-average of several instantaneous Lorentzian profiles with central frequency $\nu(t)$, given by:
\begin{equation}
    \langle P(\nu) \rangle \, = \, \frac{1}{T} \int^T_0 H \left( 1 \, + \, \left( \frac{\nu - \nu(t)}{\Gamma /2} \right)^2 \right)^{-1} dt \, ,
    \label{eq:stacked_lorentzians}
\end{equation}
where the angled brackets indicate an average over time, $H$ and $\Gamma$ are the mode height (maximum power spectral density) and linewidth, respectively, and the full period of observation is given by $T$.

\citet{chaplin_distortion_2008} also show that by assuming a simple, linear variation of the unperturbed frequency, $\nu_0$, from the start to the end of the time-series by a total frequency shift $\Delta\nu$:
\begin{equation}
    \nu(t) \, = \, \nu_0 \, +  \Delta\nu \frac{t}{T} \, ,
    \label{eq:linear_variation}
\end{equation}
the resultant profile of a $p$ mode can analytically be modelled by equation~(\ref{eq:atan_lorentzians}):
\begin{equation}
    \langle P(\nu) \rangle \, = \, \frac{H}{2\epsilon} \arctan \left( \frac{2 \epsilon}{1 - \epsilon^2 + X^2 } \right) \, ,
    \label{eq:atan_lorentzians}
\end{equation}
where $\epsilon$ and $X$ are defined in equation~\ref{eq:epsilon} and equation~\ref{eq:X}:
\begin{equation}
    \epsilon \, = \, \frac{\Delta\nu}{\Gamma} \, ;
    \label{eq:epsilon}
\end{equation}
\begin{equation}
    X \, = \, \frac{\nu - [\nu_0 + (\Delta\nu/2)]}{\Gamma /2} \, .
    \label{eq:X}
\end{equation}

As the mode linewidths are broadened by this effect, we evaluated whether our ability to resolve the true linewidth of the RM, and hence the lifetime, was affected. In order to evaluate this we computed the broadened profiles given by both equation~(\ref{eq:stacked_lorentzians}) and equation~(\ref{eq:atan_lorentzians}), and fit the model for a single Lorentzian peak, to determine whether there was a notable difference in the linewidth.

In the first instance, we computed the broadened peak using equation~(\ref{eq:stacked_lorentzians}). Over the duration of the observations, we computed the daily instantaneous profile, $P(\nu(t))$. The time-averaged profile, $ 
\langle P(\nu) \rangle$, was a weighted average of each instantaneous profile, where the weights were given by the squared-daily-SMMF, in order to allow a larger broadening contribution at times when the SMMF amplitude is higher.

In the second instance, we computed the broadened peak using equation~(\ref{eq:atan_lorentzians}). Over the duration of the observations the daily frequency shift, $\Delta\nu$, was computed. The time-averaged shift, $\Delta\nu$, was a weighted average, where again the weightings were given by the squared-daily-SMMF.

To determine the shift in the rotation rate as the active bands migrate to the solar equator, we used the model of the solar differential rotation as traced by magnetic features ($\Omega_m$) given by:
\begin{equation}
    \frac{\Omega_m}{2 \pi} \, = \, 462 - 74 \mu^2 - 53 \mu^4 \, \mathrm{nHz} \, ,
    \label{eq:diff_rot_freq}
\end{equation}
where $\mu \, = \, \cos \theta $ and $\theta$ is the co-latitude \citep{snodgrass_magnetic_1983, brown_inferring_1989}.

The time-dependence on the latitude of the active regions used the best-fitting quadratic model by \cite{li_latitude_2001-1}.

In both instances, the broadened peak was modelled as a single Lorentzian peak using equation~(\ref{eq:symm_lorentzian}). Again, we use \textsc{emcee} \citep{foreman-mackey_emcee_2013} to explore the posterior parameter space, with priors similar to the above full-fit on the relevant parameters.

\subsection{Results: Time-Averaged Broadened Profile}

\begin{figure}
\centering
\subfloat[Time-Averaged Broadened Profile \label{fig:weighted_shift}]{\includegraphics[width=0.9\columnwidth]{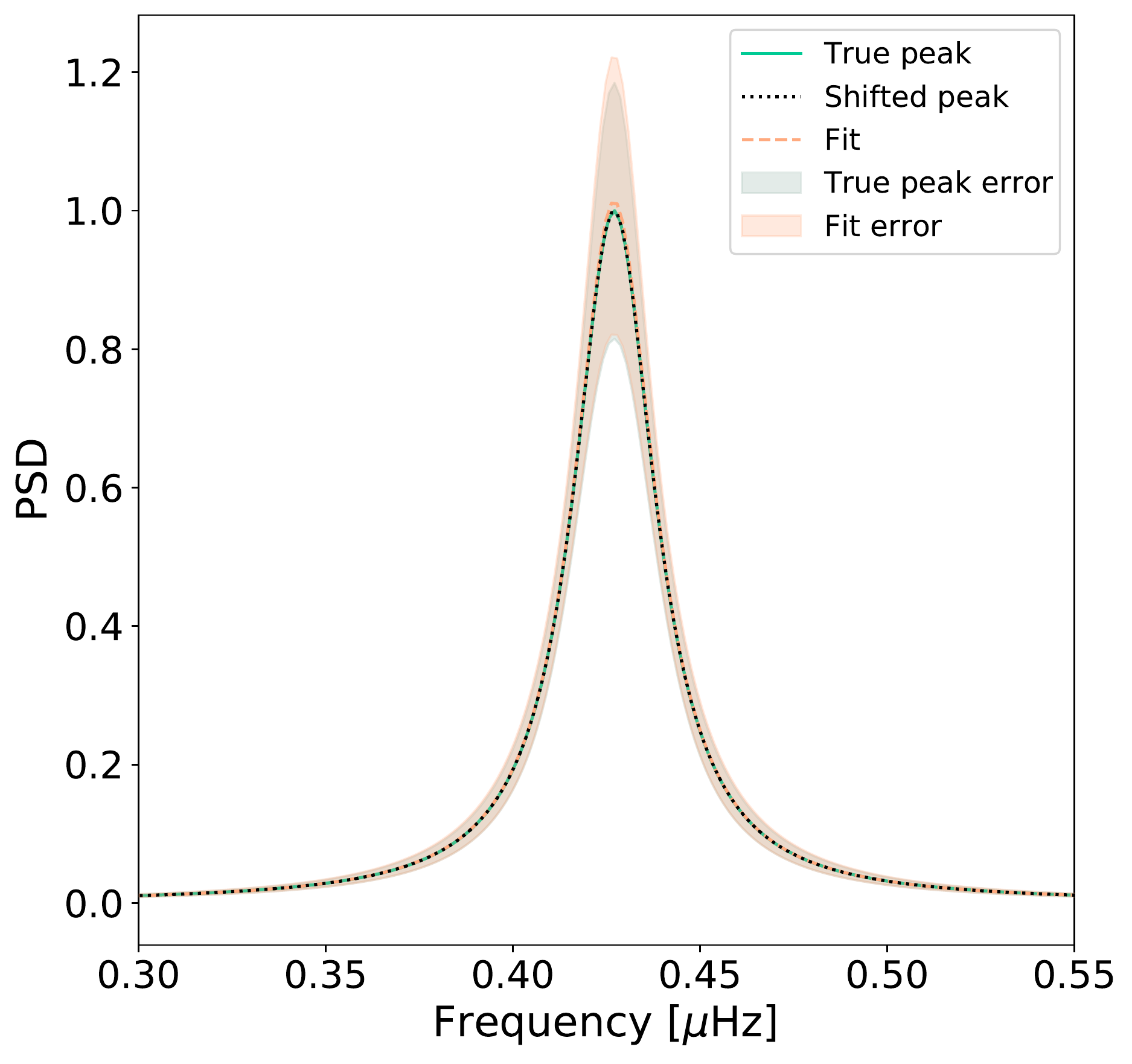}} \\ 
\subfloat[Analytically Broadened Profile \label{fig:atan_shift}]{\includegraphics[width=0.9\columnwidth]{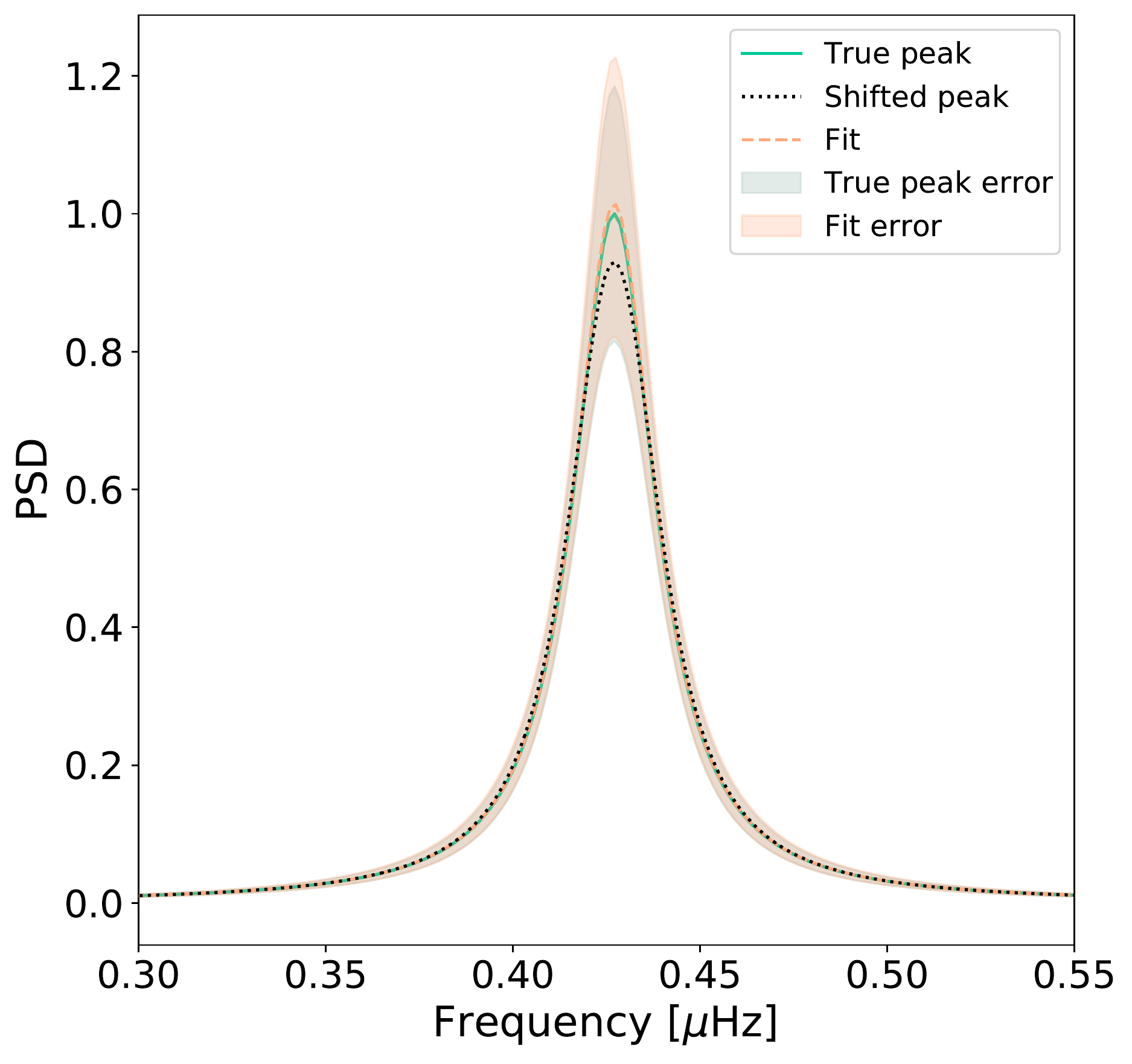}}
\caption{(a) Shows the peak distribution before and after the time-averaged broadening, and the fit to the broadened peak. (b) Shows the peak distribution before and after the analytical broadening, and the fit to the broadened peak. In both plots the broadened peaks have been shifted by the relevant frequency to overlay them on top of the true $\nu_0$ for comparison.}\label{fig:shifted_peaks}
\end{figure}

Over the entire duration of the SMMF observations, the time-averaged profile was calculated, using equation~(\ref{eq:stacked_lorentzians}), and this is shown in Fig.~\ref{fig:weighted_shift}. The broadened mode used the input parameters outlined in Table~\ref{tab:full_fit_params}, however with the background parameter set to zero.

By eye, the broadened profile does not appear to have a significantly larger linewidth. The input linewidth was $0.0264 \pm 0.0035 \, \mu\mathrm{Hz} $, and the fit to the time-averaged broadened peak produced a linewidth of $0.0262^{+0.0038}_{-0.0037} \, \mu\mathrm{Hz} $. The linewidth of the broadened peak under this method was rather unchanged from that of the true peak, and both linewidths are within uncertainties of each other.

This result shows that numerically, the mode broadening effect of differential rotation and migration does not affect our ability to resolve the linewidth of the peak, and hence the predicted lifetime of the RM component of the SMMF.

\subsection{Results: Analytically Broadened Profile}

The time-averaged frequency shift due to differential rotation was calculated, much in the same way as equation~(\ref{eq:stacked_lorentzians}), to be $\Delta\nu \, = \,0.01285 \, \mu\mathrm{Hz}$. This shift was used to generate the broadened profile using equation~(\ref{eq:atan_lorentzians}). The broadened mode distribution also used the input parameters outlined in Table~\ref{tab:full_fit_params}, however with the background parameter set to zero.

Similar to the numerically broadened peak, by eye, the analytically broadened profile does not appear to have a significantly larger linewidth (see Fig.~\ref{fig:atan_shift}). The input linewidth was $0.0264 \pm 0.0035 \, \mu\mathrm{Hz} $, and the linewidth of the analytically broadened peak from the fit was $0.0263^{+0.0038}_{-0.0037} \, \mu\mathrm{Hz} $, which was within the uncertainties of the linewidth of the input peak.

This result shows, analytically, the mode broadening effect of differential rotation and migration does not affect our ability to resolve the linewidth of the peak, and hence the lifetime of the RM component of the SMMF.

\subsection{Discussion}

Both broadening methods applied were shown to have a negligible effect on the linewidth of the profile, and our ability to resolve the true linewidth of the peak remains unaffected. This result provides confidence that the measured linewidth in Table~\ref{tab:full_fit_params} was the true linewidth of the RM peaks, providing the correct lifetime for RM component, unaffected by migration and differential rotation.

%If you want to present additional material which would interrupt the flow of the main paper, it can be placed in an Appendix which appears after the list of references.

%%%%%%%%%%%%%%%%%%%%%%%%%%%%%%%%%%%%%%%%%%%%%%%%%%

% Don't change these lines
\bsp	% typesetting comment
\label{lastpage}
\end{document}